\documentclass[aps,prl,groupedaddress]{revtex4}

\usepackage{amsfonts}
\usepackage{graphicx}

\begin{document}


\title{Green Function Method for Nonlinear Systems}


\author{Marco Frasca}
\email[]{marcofrasca@mclink.it}
\affiliation{Via Erasmo Gattamelata, 3 \\ 00176 Roma (Italy)}


\date{\today}

\begin{abstract}
We show that a Green function solution can be given for a class of 
non-homogeneous nonlinear systems having relevance in quantum field theory. This in
turn means that a quantum field theory in the strong coupling limit can be formulated
and the spectrum obtained.  
\end{abstract}


\maketitle


It is common wisdom that Green function methods are typical of linear systems. There
is no way to write down a solution to a non-homogeneous non-linear problem in a closed
analytical form using the solution of the correspondent equation for the Green function.

The reason for this is quite easy to understand as the Green function method relies on
an integral formula that implies the validity of the superposition principle. This point
is particularly relevant in quantum field theory where strong coupling generally means 
to take the non-linear part of classical equations of the field theory without possibility
to rely on the superposition principle and weak perturbation theory.

Recently we showed how a perturbation theory can be obtained for strongly perturbed
field theories \cite{fra1,fra2,fra3,fra4}. The main result can be summed up as to
have proved that a gradient expansion is a strong perturbation theory. This in turn
implies that the leading order is generally a classical equation in the form
\begin{equation}
\label{eq:base}
    \ddot\phi(t)+P(\phi(t))=0
\end{equation}
being $P(\phi(t))$ a generic non-linear function of the field $\phi(t)$. A Green function
can anyway be defined in this case by looking for a solution of the equation
\begin{equation}
    \ddot\phi(t)+P(\phi(t))=A\delta(t),
\end{equation}
being $A$ a constant scale factor, but there is no way to attach a meaning to a solution
of the equation
\begin{equation}
     \ddot\phi(t)+P(\phi(t))=j(t),
\end{equation} 
being $j(t)$ a given source, in the well-known form
\begin{equation}
\label{eq:gf}
     \phi(t)=\int_{-\infty}^{+\infty}G(t-t')j(t')dt'.
\end{equation} 
Anyhow, using a pure numerical approach, we are going to show that although eq.(\ref{eq:gf})
is not exactly applicable, an approximate solution is given in this case by
\begin{equation}
\label{eq:main}
     \phi(t)\approx\int_0^tG(t-t')j(t').
\end{equation}
On a mathematical standpoint this appears as the leading order of an expansion to hold
in the strong coupling limit, when we expand with respect to a given large parameter
in eq.(\ref{eq:base}). Then, when we interpret the Green function in quantum field
theory as containing all the spectrum, we are able to obtain it in the limit
of a strong coupling limit holding at the leading order. Indeed we used this approach to
obtain the spectrum in the strong coupling limit of a $\phi^4$ theory\cite{fra3}. 
Here we will support our finding by applying this approach also to the pendulum equation
\begin{equation}
    \ddot\vartheta(t)+\sin(\vartheta(t))=0
\end{equation}
that appears to be the leading order of a gradient expansion of the sine-Gordon model in
quantum field theory. We will obtain the spectrum for this field theory in the strong coupling limit, 
that is, when the non-linear term is taken to be largely increasing.

\begin{figure}[tbp]
\begin{center}
\includegraphics[angle=0,width=240pt]{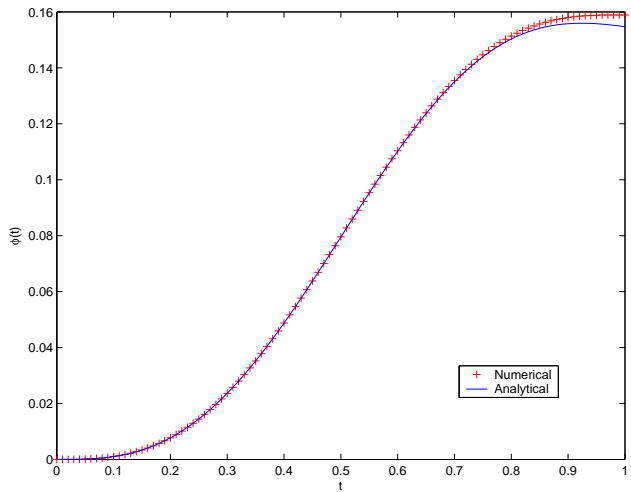}
\caption{\label{fig:fig1} Agreement between numerical and analytical solutions for
$\phi^4$ case with a sine forcing.}
\end{center}
\end{figure}

\begin{figure}[tbp]
\begin{center}
\includegraphics[angle=0,width=240pt]{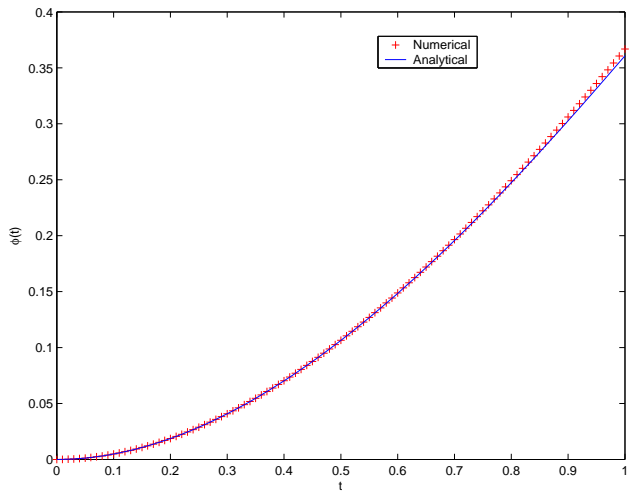}
\caption{\label{fig:fig2} Agreement between numerical and analytical solutions for
$\phi^4$ case with an exponential forcing.}
\end{center}
\end{figure}

Firstly, let us review the situation of a $\phi^4$ case. A gradient expansion gives at
the leading order the classical equation
\begin{equation}
    \ddot\phi(t)+\phi^3(t)=0.
\end{equation}
We are interested to the numerical solution of the non-homogeneous equation
\begin{equation}
    \ddot\phi(t)+\phi^3=\sin(2\pi t).
\end{equation}
The solution of the Green function equation
\begin{equation}
\label{eq:phi4}
    \ddot G(t)+G(t)^3=\delta(t)
\end{equation} 
is
\begin{equation}
    G(t)=\theta(t)2^{\frac{1}{4}}
	{\rm sn}\left[\left(\frac{1}{2}\right)^{\frac{1}{4}}t,i\right]
\end{equation}
being ${\rm sn}$ the Jacobi snoidal function and $\theta(t)$ the Heaviside function. So, we
compare our numerical solution with the analytical formula
\begin{equation}
    \phi(t)\approx 2^{\frac{1}{4}}\int_0^t
	{\rm sn}\left[\left(\frac{1}{2}\right)^{\frac{1}{4}}(t-t'),i\right]
	\sin(2\pi t')dt'.
\end{equation}
The results are given in fig.(\ref{fig:fig1}). The agreement appears to be perfect till the
end of the integration interval where both curves start to differ. This is a typical behavior
of a term of a perturbation series.
We have also tried to change the driving term. We have considered the case $j(t)=e^{-t}$
and the result is given in fig.(\ref{fig:fig2}). Again the agreement is practically
perfect for a large part of the integration interval.

\begin{figure}[tbp]
\begin{center}
\includegraphics[angle=0,width=240pt]{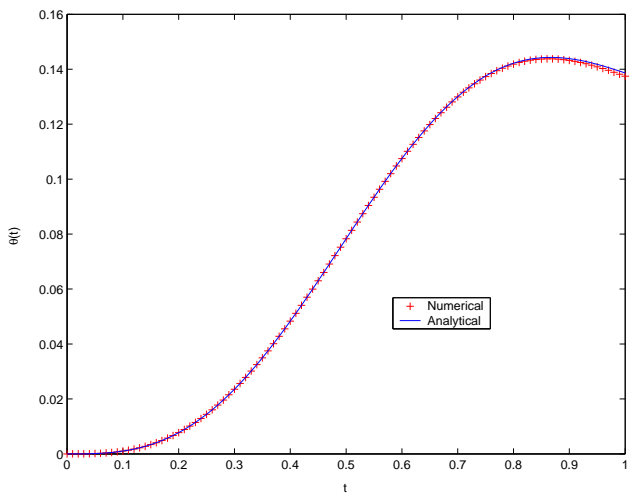}
\caption{\label{fig:fig3} Agreement between numerical and analytical solutions for
the pendulum case with a sine forcing.}
\end{center}
\end{figure}

The Sine-Gordon equation has an exterminate literature. The quantum model in 1+1 dimensions
was firstly considered by Coleman \cite{cole} obtaining the classical result that this
model is equivalent to the  massive Thirring model. Through a gradient expansion at the
leading order, a forced pendulum can be obtained from the Sine-Gordon equation 
written in the form
\begin{equation}
\label{eq:sg}
    \ddot\vartheta-\nabla^2\vartheta+\sin(\vartheta)=j
\end{equation}
A forced Sine-Gordon equation has been considered
in \cite{fio1,fio2}. Our case applies when the non-linear term is very large. In this case
one can show that a gradient expansion is obtained \cite{fra1,fra2,fra3,fra4}.
We will give a proof of this below. So, we have
to solve the forced pendulum equation
\begin{equation}
\label{eq:pend}
    \ddot\vartheta(t)+\sin(\vartheta(t))=j(t).
\end{equation}
In order to apply our approach we consider the following Green function
\begin{equation}
    G(t)=2\theta(t){\rm am}\left[\frac{1}{\sqrt{2}}t,\sqrt{2}\right],
\end{equation}
being ${\rm am}$ the Jacobi amplitude, that solves the equation
\begin{equation}
\label{eq:gp}
    \ddot G(t)+\sin(G(t))=\sqrt{2}\delta(t).
\end{equation}
Then, with a forcing sine function we get
\begin{equation}
    \theta(t)\approx \frac{2}{\sqrt{2}}\int_0^t{\rm am}\left[\frac{1}{\sqrt{2}}(t-t'),\sqrt{2}\right]\sin(2\pi t')dt'.
\end{equation}
having rescaled by $\sqrt{2}$ to take into account the definition eq.(\ref{eq:gp})
As shown in Fig.(\ref{fig:fig3}), the agreement between numerical and analytical results
is satisfactory.

The relevance of this result relies on the fact that is quite easy to write down
a generating functional for a leading order quantum field theory. Indeed, the classical
theory is very similar to a free linear theory. We have already used this fact in Ref.\cite{fra3}
to obtain the spectrum and the propagator of a $\lambda\phi^4$ theory in the limit of
very large $\lambda$. We have shown that the generating functional can be written down as
\begin{equation}
    Z[j]=\exp\left[\frac{i}{2}\int d^Dxd^Dx'j(x)\Delta(x-x')j(x')\right]
\end{equation}
being $\Delta(x-x')$ the Feynman propagator. In our case this propagator can be easily
obtained using the fact that we have two solutions to eq.(\ref{eq:phi4}) 
and eq.(\ref{eq:gp}) obtained just
by the change $t\leftrightarrow -t$. In this way we are able to recover our solution
eq.(\ref{eq:main}) as it should be with the proper time evolution. This can be easily
accomplished when the Feynman propagator is written down as \cite{fra3}
\begin{equation}
    \Delta(x)=-i\delta^{(D-1)}(x)[G(t)+G(-t)]
\end{equation}
to take into account the given time ordering. So, let us consider a source $j(t)$ starting at t=0 at
increasing time. One has
\begin{equation}
    \frac{\delta Z[j]}{\delta j(x)}=\phi(x)=\int_0^t \tilde G(t-t')j(t')dt'
\end{equation}
being $\tilde G$ the Green function with the Heaviside function removed and we have integrated
the Dirac distribution on the spatial coordinates.

In order to write down both the spectrum and the propagator 
for our case we use the following results \cite{gr}:
\begin{equation}
{\rm sn}(u,i)=\frac{2\pi}{K(i)}\sum_{n=0}^\infty\frac{(-1)^ne^{-(n+\frac{1}{2})\pi}}{1+e^{-(2n+1)\pi}}
    \sin\left[(2n+1)\frac{\pi u}{2K(i)}\right]
\end{equation}
being $K(i)=\int_0^{\frac{\pi}{2}}\frac{d\theta}{\sqrt{1+\sin^2\theta}}\approx 1.3111028777$
a constant. Similarly,
\begin{equation}
{\rm am}(u,\sqrt{2})=\frac{\pi u(1+i)}{2K(i)}+
    \sum_{n=1}^\infty\frac{1}{n}\frac{(-i)^ne^{-n\frac{\pi}{2}}}{1+(-1)^ne^{-n\pi}}
    \sin\left[\frac{n\pi u(1+i)}{2K(i)}\right].
\end{equation}
Our aim is to get the time dependent part of the propagator in the form
\begin{equation}
    \tilde G(t)=\sum_{n=-\infty}^\infty A_n e^{-iE_nt}
\end{equation}
being $A_n$ constant amplitudes and $E_n$ the spectrum of the given quantum field theory.
For the $\phi^4$ theory one has easily
\begin{equation}
    \tilde G(t)=\sum_{n=0}^\infty\frac{\pi}{iK(i)}
	\frac{(-1)^ne^{-(n+\frac{1}{2})\pi}}{1+e^{-(2n+1)\pi}}\left[e^{i(2n+1)\frac{\pi}{2^\frac{5}{4}K(i)}t}
	-e^{-i(2n+1)\frac{\pi}{2^\frac{5}{4}K(i)}t}\right]
\end{equation}
and we are able to read out the spectrum being given by that of a harmonic oscillator
\begin{equation}
    E_n=(2n+1)\frac{\pi u}{2^\frac{5}{4}K(i)}
\end{equation}
for the particle content of the theory in the strong coupling limit \cite{fra3}. Similarly
one has
\begin{equation}
    A_n=\frac{\pi}{iK(i)}\frac{(-1)^ne^{-(n+\frac{1}{2})\pi}}{1+e^{-(2n+1)\pi}}
\end{equation}
and its complex conjugate. For the Sine-Gordon case we arrive at a rather unexpected
result that the quantum theory in a strong coupling limit (increasingly large mass) has
complex energy eigenvalues. Indeed, one has
\begin{equation}
    \tilde G(t)=\frac{\pi(1+i)t}{\sqrt{2}K(i)}+
    \sum_{n=1}^\infty\frac{1}{n}\frac{(-i)^ne^{-n\frac{\pi}{2}}}{1+(-1)^ne^{-n\pi}}
    \sin\left[\frac{n\pi(1+i)t}{2\sqrt{2}K(i)}\right]
\end{equation}
and this gives the result
\begin{equation}
\tilde G(t)=\frac{\pi(1+i)t}{\sqrt{2}K(i)}+
    \sum_{n=1}^\infty\frac{1}{in}\frac{(-i)^ne^{-n\frac{\pi}{2}}}{1+(-1)^ne^{-n\pi}}
	\left[e^{i\frac{n\pi t}{2\sqrt{2}K(i)}}e^{\frac{-n\pi t}{2\sqrt{2}K(i)}}
	-e^{-i\frac{n\pi t}{2\sqrt{2}K(i)}}e^{\frac{n\pi t}{2\sqrt{2}K(i)}}\right]
\end{equation} 
that means
\begin{equation}
    E_n=\frac{n\pi}{\sqrt{2}K(\sqrt{2})}
\end{equation}
being $K(\sqrt{2})=K(i)(1-i)$ and
$n$ starting at $n=1$. We note also the presence of a zero mode. It
should be noticed that the Green function of the Sine-Gordon model is bounded in the
limit $t\rightarrow\infty$ making sense of complex eigenvalues, but each single mode
is unstable with respect to time evolution. This in turn implies that physical
results, obtained using the full Green function, are also finite. 
We note that this situation has been studied in Bose-Einstein
condensates \cite{mine}.

In order to complete our analysis, we show how a gradient expansion emerges as a strong coupling
expansion for the particular case of the Sine-Gordon model. This analysis also applies
to the $\phi^4$ model as already shown \cite{fra1,fra2,fra3,fra4}. We reintroduce the parameters
into eq.(\ref{eq:sg}) to write
\begin{equation}
    \ddot\vartheta-\nabla^2\vartheta+\frac{\alpha}{\beta}\sin(\beta\vartheta)=0
\end{equation}
being $\alpha$ a squared mass term and we take $\beta=1$ preserving consistency in agreement
to the analysis given by Coleman \cite{cole}. A gradient expansion is recovered by
rescaling time as $t\rightarrow\sqrt{\alpha} t$ and putting
\begin{equation}
     \vartheta = \vartheta_0+\frac{1}{\alpha}\vartheta_1+\frac{1}{\alpha^2}\vartheta_2+\ldots
\end{equation}
showing that we do have a strong coupling expansion that holds in the limit $\alpha\rightarrow\infty$. 
Indeed, we get the following non trivial set of equations
\begin{eqnarray}
     \ddot\vartheta_0+\sin(\vartheta_0)&=& 0 \\ \nonumber
	 \ddot\vartheta_1+\cos(\vartheta_0)\vartheta_1&=&\nabla^2\vartheta_0 \\ \nonumber
	 \ddot\vartheta_2+\cos(\vartheta_0)\vartheta_2&=&\nabla^2\vartheta_1 +\frac{1}{2}\sin(\vartheta_0)\vartheta_1^2\\ \nonumber
	 &\vdots&.
\end{eqnarray}
Adding a source turns us back to eq.(\ref{eq:pend}) for the leading order. This complete
our proof.

Although we have shown, through numerical analysis, that for some nonlinear models the
Green function appears to work, only an analytical proof could move our results from a
conjectural state to a fully proved theorem. Notwithstanding this aspect of our approach,
the results are really unexpected as it is a general wisdom to assume that only linear
problems could be treated through Green function methods, as happens also to the leading order
of a weak perturbation theory. The possibility to get a full spectrum of a quantum
field theory, in some limit, is a very relevant by-product of this method and the
nature of a gradient expansion, both in the classical and quantum limits, is elucidated
appearing as a natural strong coupling expansion.


\end{document}